\documentclass[aps,prl,preprint,showpacs,groupedaddress]{revtex4}
\usepackage {amssymb}
\usepackage {amsmath}
\usepackage {graphicx}
\usepackage {longtable}

\begin{document}
\title{Effect of magnetic ordering on the electronic structure of metals}

\author{Fedor V.Prigara}
\affiliation{Institute of Physics and Technology,
Russian Academy of Sciences,\\
21 Universitetskaya, Yaroslavl 150007, Russia}
\email{fprigara@recnti.uniyar.ac.ru}

\date{\today}

\begin{abstract}

It is shown that ferromagnetic ordering in metals is associated with an
opening of the energy pseudogap. This energy pseudogap belongs to a d-band
type in d-metals and to an sp-band type in f-metals. A relation between the
magnetic energy and the Curie temperature is obtained. Effect of magnetic
ordering on the temperature dependence of the electrical resistivity is
considered.

\end{abstract}


\maketitle

Antiferromagnetic ordering in metals is associated with an opening of the
energy pseudogap at the Fermi level [1]. The magnitude of the
antiferromagnetic pseudogap is proportional to the Neel temperature [2,3].
Here we show that ferromagnetic ordering in metals is also associated with
an opening of the energy pseudogap at the Fermi level. A relation between
the magnitude of the ferromagnetic pseudogap and the Curie temperature is,
however, different in d-metals and f-metals. The ferromagnetic pseudogap in
d-metals belongs to a d-band type [3], whereas the ferromagnetic pseudogap
in f-metals belongs to an sp-band type, so that a relation between the
magnitude of the energy pseudogap and the magnetic ordering temperature in
f-metals is similar to those for antiferromagnetic ordering.

The full magnitude $\Delta _{AFM} \left( {0} \right)$ of the
antiferromagnetic pseudogap at zero temperature ($T = 0K$) is related to the
Neel temperature $T_{N} $ by the formula [3]

\begin{equation}
\label{eq1}
\Delta _{AFM} \left( {0} \right) = \alpha k_{B} T_{N} ,
\end{equation}

\noindent
where $k_{B} $ is the Boltzmann constant, and $\alpha = 18$ is a constant.

The equation (\ref{eq1}) means that the energy $E_{AFM} $ of an elementary
antiferromagnetic excitation [4] is equal to the magnitude $\Delta _{AFM} $
of the antiferromagnetic pseudogap, so that the pressure dependence of the
magnitude of the antiferromagnetic pseudogap in the high-pressure region is
given by the equation

\begin{equation}
\label{eq2}
\Delta _{AFM} \left( {P} \right) = \Delta _{AFM} \left( {0} \right) - \alpha
_{P} P/n_{0} ,
\end{equation}

\noindent
where \textit{P} is the pressure, $n_{0} \approx 1.1 \times 10^{22}cm^{ -
3}$ is a constant which has an order of the number density of atoms in the
crystalline state ($a_{0} = n_{0}^{ - 1/3} \approx 0.45nm$has an order of
the lattice parameter), and $\alpha _{P} $ is the atomic relaxation
constant.

The equations (\ref{eq1}) and (\ref{eq2}) give the pressure dependence of the Neel
temperature in the high-pressure region in the form

\begin{equation}
\label{eq3}
k_{B} T_{N} \left( {P} \right) = k_{B} T_{N} \left( {0} \right) -
\frac{{\alpha _{P}} }{{\alpha} }\frac{{P}}{{n_{0}} }.
\end{equation}

A comparison of the equation (\ref{eq3}) with the experimental data for $CeIn_{3} $
[5[ gives a value of the atomic relaxation constant for antiferromagnetic
ordering $\alpha _{P} = 2$. The atomic relaxation constant for
antiferromagnetic ordering is equal to the atomic relaxation constant for
the metal-insulator transition [6], in agreement with those fact that the
antiferromagnetic transition can coincide with the metal-insulator
transition, for example, in the pyrochlore iridate $Eu_{2} Ir_{2} O_{7} $
[7] and also in underdoped cuprate high-temperature superconductors [8].

If the antiferromagnetic transition occurs in the insulating phase, for
example, in NiO, it causes a splitting of the valence band. A split-off band
is formed mainly by the d-states, and the upper band is formed mainly by the
sp-states. The magnitude $\Delta $ of a splitting at the top of the valence
band at zero temperature ($T = 0K$) is given by the formula

\begin{equation}
\label{eq4}
\Delta \left( {0} \right) = \alpha _{G} \alpha k_{B} T_{N} ,
\end{equation}

\noindent
where $\alpha _{G} = 3/8$ is the gap constant.

A valence band splitting is temperature dependent and vanishes at the Neel
temperature. The Neel temperature in NiO is $T_{N} = 530K$ [9], and the
equation (\ref{eq4}) gives the magnitude of a valence band splitting at zero
temperature at a level of $\Delta \left( {0} \right) = 0.31eV$. The
experimental value of a valence band splitting in NiO from optical
absorption measurements is $\Delta = 0.24eV$ at $T = 300K$ [10].

A splitting of the valence band in NiO is caused by a rhombohedral
distortion of a rocksalt type crystal structure below the Neel temperature.
The rhombohedral angle is about $60^{ \circ} 04^{'}$ at room temperature
[11], so that a ferroelastic distortion associated with antiferromagnetic
ordering in NiO is small. It corresponds to a relative contraction of the
lattice at zero temperature ($T = 0K$). A similar rhombohedral distortion of
a rocksalt type crystal structure is present in MnO below the Neel
temperature $T_{N} \cong 120K$.

Since the density of states in the d-band is much higher than those in the
sp-band, a splitting of the valence band can look like an increase of the
bandgap width $E_{g} \left( {0} \right)$ at zero temperature ($T = 0K$) in
optical absorption measurements by a value of

\begin{equation}
\label{eq5}
\Delta E_{g} \left( {0} \right) = \frac{{1}}{{2}}\Delta \left( {0} \right) =
\frac{{1}}{{2}}\alpha _{G} \alpha k_{B} T_{N} .
\end{equation}

Such is the case in $BiFeO_{3} $ [12], where the Neel temperature is $T_{N}
= 640K$, and the equation (\ref{eq5}) gives $\Delta E_{g} \left( {0} \right) =
0.18eV$.

There is, however, a similar splitting if the valence band in Ge which is
not associated with antiferromagnetic ordering. In this case, the magnitude
of a splitting of the valence band at zero temperature ($T = 0K$) is related
to the metal-insulator transition temperature $T_{MI} $ by the formula

\begin{equation}
\label{eq6}
\Delta \left( {0} \right) = \alpha _{G} \alpha k_{B} T_{MI} ,
\end{equation}

\noindent
with the same gap constant $\alpha _{G} = 3/8$.

There seems to be a contribution of the 4d-orbitals to the wave functions of
electrons in Ge, in view of the relation (\ref{eq6}). In Si, this effect is absent.

A splitting of the valence band in Ge is caused by a rhombohedral distortion
of a diamond-type crystal structure associated with the metal-insulator
transition [6]. A rhombohedral ferroelastic distortion in Ge produces also a
large anisotropy of the effective mass of electrons. The dispersion of the
conduction band in Ge along the [111] direction is weak due to a weaker
overlap of the 4sp-orbitals along this direction. A feroelastic distortion
associated with the metal-insulator transition corresponds to a relative
expansion of the lattice at zero temperature ($T = 0K$).

The magnetic energy $E_{M} $ in a metal is related to the full magnitude
$\Delta \left( {0} \right)$ of the energy pseudogap associated with magnetic
ordering at zero temperature ($T = 0K$) by the equation similar to a
relation between the condensation energy in a superconductor and the
magnitude of the superconducting gap [13]

\begin{equation}
\label{eq7}
E_{M} \cong \frac{{1}}{{2}}N\left( {E_{F}}  \right)\left( {\Delta \left( {0}
\right)/2} \right)^{2} = \frac{{1}}{{8}}N\left( {E_{F}}  \right)\Delta
^{2}\left( {0} \right).
\end{equation}

Here $N\left( {E_{F}}  \right)$ is the density of states at the Fermi level
which can be determined from the electronic specific heat coefficient
$\gamma $ [14],

\begin{equation}
\label{eq8}
\gamma = \frac{{2\pi ^{2}}}{{3}}N\left( {E_{F}}  \right)k_{B}^{2} .
\end{equation}

Both in the equation (\ref{eq7}) and in the equation (\ref{eq8}) $N\left( {E_{F}}  \right)$
is a real density of states at the Fermi level, with account for many-body
effects.

From the equations (\ref{eq1}), (\ref{eq7}), and (\ref{eq8}), we find the magnetic energy in
antiferromagnetic metals in the form

\begin{equation}
\label{eq9}
E_{M} \cong \frac{{3}}{{16\pi ^{2}}}\gamma \left( {\alpha T_{N}}
\right)^{2} \cong 6.1\gamma T_{N}^{2} .
\end{equation}

For antiferromagnetic ordering in Mn with the Neel temperature $T_{N} = 95K$
[15] and the electronic specific heat coefficient $\gamma = 10.6mJmol^{ -
1}K^{ - 2}$, the equation (\ref{eq9}) gives the magnetic energy at a level of $E_{M}
\cong 0.58kJmol^{ - 1}$, or $E_{M} \cong 0.73k_{B} T_{N} $ per atom.

In the case of ferromagnetic ordering in a metal, a relation between the
full magnitude\textbf{} $\Delta _{FM} \left( {0} \right)$ of the energy
pseudogap at zero temperature \textbf{(}$T = 0K$\textbf{)} and the
Curie\textbf{} temperature\textbf{} $T_{c} $ should be modified with respect
to the equation (\ref{eq1}) as follows

\begin{equation}
\label{eq10}
\Delta _{FM} \left( {0} \right) = \alpha _{G} \alpha k_{B} T_{c} ,
\end{equation}

\noindent
where $\alpha _{G} $ is the gap constant.

In this case, the equations (\ref{eq7}), (\ref{eq8}), and (\ref{eq10}) give the magnetic energy in
the form

\begin{equation}
\label{eq11}
E_{M} \cong \frac{{3}}{{16\pi ^{2}}}\left( {\alpha _{G} \alpha}
\right)^{2}\gamma T_{c}^{2} .
\end{equation}

For ferromagnetic ordering in Fe with the Curie temperature $T_{c} = 1043K$
and the electronic specific heat coefficient $\gamma = 5.02mJmol^{ - 1}K^{ -
2}$, a comparison of the equation (\ref{eq11}) with an experimental value of the
magnetic energy in Fe determined from the specific heat data [15] gives a
value of the gap constant $\alpha _{G} = 3/8$. The magnitude of the
ferromagnetic pseudogap in Fe, according to the equation (\ref{eq10}), is $\Delta
_{FM} \left( {0} \right) = 0.61eV$. The magnetic energy in Fe is $E_{M}
\cong 4.8kJmol^{ - 1}$, or $E_{M} \cong 0.56k_{B} T_{c} $ per atom.

For ferromagnetic ordering in Co with the Curie temperature $T_{c} = 1394K$
and the electronic specific heat coefficient $\gamma = 5.02mJmol^{ - 1}K^{ -
2}$, the equation (\ref{eq11}) with the gap constant $\alpha _{G} = 3/8$ gives the
magnetic energy $E_{M} \cong 7.8kJmol^{ - 1}$, or $E_{M} \cong 0.68k_{B}
T_{c} $ per atom.

Ferromagnetic transitions in d-metals, in view of the equation (\ref{eq10}), belong
to a d-band type, according to a classification introduced in Ref. 3.
Antiferromagnetic transitions in metals, in view of the equation (\ref{eq1}), belong
to an sp-band type.

For ferromagnetic ordering in Gd with the Curie temperature $T_{c} = 290K$
and the electronic specific heat coefficient $\gamma \cong 8mJmol^{ - 1}K^{
- 2}$, a comparison of the equation (\ref{eq11}) with an experimental value of the
magnetic energy determined from the specific heat data [15] gives a value of
the gap constant $\alpha _{G} = 1$. Therefore, the magnitude $\Delta _{FM}
\left( {0} \right)$ of the ferromagnetic pseudogap at zero temperature in
f-metals is related to the Curie temperature $T_{c} $ by the equation
similar to the equation (\ref{eq1}),

\begin{equation}
\label{eq12}
\Delta _{FM} \left( {0} \right) = \alpha k_{B} T_{c} .
\end{equation}

The magnitude of the ferromagnetic gap in Gd, according to the equation
(\ref{eq12}), is $\Delta _{FM} \left( {0} \right) = 0.45eV$, the magnetic energy in
Gd is $E_{M} \cong 4kJmol^{ - 1}$, or $E_{M} \cong 1.7k_{B} T_{c} $ per
atom. The magnitude of the ferromagnetic pseudogap in Gd is equal to the
magnitude of the energy pseudogap associated with the hcp-bcc transition at
$T_{s} = 1533K$ which belongs to a d-band type with the atomic relaxation
constant $\alpha _{P} = 3/16$ [3]. Ferromagnetic transitions in f-metals
belong to an sp-band type, in agreement with a character of the exchange
interaction via the conduction electrons.

There is an experimental evidence for an opening of the energy pseudogap in
Gd below the curie temperature $T_{c} = 290K$, and also in Dy below the
magnetic ordering temperature $\theta _{2} = 179K$ from optical reflection
measurements [16].

For magnetic ordering of the Yb moments below $T_{M} = 5K$ in $Yb_{2}
Co_{12} P_{7} $ with the electronic specific heat coefficient $\gamma =
77mJmol\left( {Yb} \right)^{ - 1}K^{ - 2}$ [17], the equation (\ref{eq9}) gives the
magnetic energy $E_{M} \cong 12Jmol\left( {Yb} \right)^{ - 1}$, or $E_{M}
\cong 0.3k_{B} T_{M} $ per atom. The experimental value of the magnetic
energy associated with the magnetic transition at $T_{M} = 5K$ in $Yb_{2}
Co_{12} P_{7} $, which can be determined from the specific heat data, is
about $E_{M} \cong 10Jmol\left( {Yb} \right)^{ - 1}$ [17].

There is an increase in the slope, $d\rho /dT$, of the temperature
dependence of the resistivity of a metal below the ferromagnetic ordering
temperature [15,17], which can be attributed to a decrease in the effective
number density of charge carriers due to an opening of the energy pseudogap.

In the free electron model, the electrical resistivity $\rho $ of a metal is
determined by the formula

\begin{equation}
\label{eq13}
\rho = \frac{{p_{F}} }{{ne^{2}}}\frac{{1}}{{l}}.
\end{equation}

Here \textit{n} is the number density of electrons, \textit{e} is the charge
of an electron, \textit{l} is the mean free path of electrons, and $p_{F} $
is the Fermi momentum given by the equation

\begin{equation}
\label{eq14}
p_{F} = \hbar \left( {3\pi ^{2}n} \right)^{1/3},
\end{equation}

\noindent
where $\hbar $ is the Planck constant.

Below the low-temperature ferroelastic transition [18] at

\begin{equation}
\label{eq15}
T_{f} \cong \theta _{D} /\alpha ,
\end{equation}

\noindent
where $\theta _{D} $ is the Debye temperature, the mean free path of
electrons is equal to the mean size of ferroelastic subdomains (subgrains),
so that the resistivity of a metal is approximately constant. For example,
in Cu the Debye temperature is $\theta _{D} = 310K$, so that the
ferroelastic transition temperature given by the equation (\ref{eq15}) is $T_{f}
\cong 17K$. The size of the ferroelastic domains can be determined from the
thermal conductivity data [18] and is about $120\mu m$. For the number
density of electrons $n = 8.4 \times 10^{22}cm^{ - 3}$ and the mean free
path of electrons $l = 10\mu m$, the equations (\ref{eq13}) and (\ref{eq14}) give the
residual resistivity of $\rho _{0} = 0.0064\mu \Omega cm$. The experimental
value of the residual resistivity in Cu depends on the sample and is about
$\rho _{0} = 0.004\mu \Omega cm$[19], which corresponds to the mean free
path of electrons $l = 16\mu m$.

In dilute alloys of noble metals (Cu, Ag, and Au) containing a small amount
of magnetic impurities (Cr, Mn, Fe), there is an additional scattering of
electrons below the ferroelastic transition temperature $T_{f} $ by
ferroelastic domains boundaries which coincide with magnetic domain walls.
The direction of a ferroelastic distortion in dilute alloys is an easy axis
for magnetic moments. In dilute alloys with an fcc crystal structure, a
ferroelastic distortion is presumably rhombohedral, directed perpendicular
to the close packed planes.

The energy of an elementary ferroelastic excitation [18] corresponds to the
energy of transverse optical phonons propagating along the direction of a
ferroelastic distortion.

In this case, the equation (\ref{eq13}) should be modified as follows

\begin{equation}
\label{eq16}
\rho = \frac{{p_{F}} }{{ne^{2}}}\left( {\frac{{1}}{{l}} + \frac{{1}}{{l_{1}
}}} \right).
\end{equation}

Here $l_{1} $ is the mean free path of electrons with respect to the
magnetic scattering and is equal to the mean size of ferroelastic domains.

There is a minimum of the resistivity at the ferroelastic transition
temperature $T_{f} $, where the resistivity is determined by the equation
(\ref{eq13}). A relative increase of the resistivity at zero temperature
\textbf{(}$T = 0K$\textbf{)} is given by the formula

\begin{equation}
\label{eq17}
\frac{{\Delta \rho} }{{\rho _{0}} } = \frac{{l}}{{l_{1}} }.
\end{equation}

Since $\Delta \rho /\rho _{0} \cong 0.08$ [19], the ratio of the mean
ferroelastic domain size $l_{1} $ to the mean ferroelastic subdomain size
\textit{l} is approximately constant, $l_{1} \cong 12l$. The size of
ferroelastic domains and subdomains decreases with increasing concentration
of impurity atoms. For $Cu_{1 - x} Fe_{x} $ with $x = 2.2 \times 10^{ - 5}$,
the residual resistivity is $\rho _{0} = 0.031\mu \Omega cm$[19], so that
the mean size of ferroelastic subdomains, according to the equation (\ref{eq13}), is
about $l \cong 2\mu m$, and the mean size of ferroelastic domains is $l_{1}
\cong 24\mu m$.

There is a critical concentration of impurity atoms above which the magnetic
scattering occurs. For $Cu_{1 - x} Fe_{x} $, a critical concentration is
$x_{0} = 2 \times 10^{ - 5}$ and corresponds to a mean distance between the
Fe atoms of $d = \left( {nx_{0}}  \right)^{ - 1/3} \cong 8nm$, which has an
order of the radius of the atomic relaxation region [4].

A second-order phase transition in $URu_{2} Si_{2} $ at $T_{h} = 17.5K$ is a
low-temperature ferroelastic transition. There are orthorhombic ferroelastic
domains below the transition temperature [20]. The size of domains has an
order of tens micrometers, similarly to the case of $Cu_{1 - x} Fe_{x} $.
There seems to be a charge-ordering (charge-density-wave) transition
coinciding with a ferroelastic transition in $URu_{2} Si_{2} $, $T_{s} =
T_{h} $, since most of charge carriers disappear below the transition
temperature. The magnitude of the charge gap is given by the equation [18]

\begin{equation}
\label{eq18}
\Delta _{ch} = \alpha k_{B} T_{s} ,
\end{equation}

\noindent
which gives $\Delta _{ch} = 0.027eV$.

The pressure-dependent optical conductivity spectra of $CeIn_{3} $ [5] show
that there is a charge-ordering (charge-density-wave) transition in this
intermetallic compound coinciding with the antiferromagnetic transition at
ambient pressure, $T_{s} = T_{N} \cong 10K$. The charge-ordering transition
temperature $T_{s} $ slightly increases with increasing pressure, since the
magnitude $\Delta _{ch} $ of the charge gap related to $T_{s} $ by the
equation (\ref{eq18}) increases in the low-pressure region from $\Delta _{ch} \cong
17.5meV$ to $18.5meV$.

There is a decrease in the resistivity of $Yb_{2} Co_{12} P_{7} $ below the
magnetic ordering temperature $T_{M} = 5K$ [17], which is lower than the
ferroelastic transition temperature determined by the equation (\ref{eq15}), due to
a change in the size of ferroelastic domains caused by magnetic ordering.
$Yb_{2} Co_{12} P_{7} $ has a hexagonal crystal structure. A ferroelastic
distortion below the ferroelastic transition temperature is presumably
orthorhombic, similarly to the case of a structural transition in $BaVS_{3}
$ [21]. There is a further monoclinic lattice distortion below the magnetic
ordering temperature $T_{M} = 5K$.

To summerize, we show that ferromagnetic ordering in metals is
associated with an opening of the energy pseudogap, similarly to
the case of antiferromagnetic ordering. We obtain a relation
between the magnitude of the energy pseudogap and the Curie
temperature, and also a relation between the magnetic energy and
the magnetic ordering temperature. Ferromagnetic transitions in
d-metals belong to a d-band type, and ferromagnetic transitions in
f-metals belong to an sp-band type. We consider an effect of
magnetic ordering on the temperature dependence of the electrical
resistivity of a metal. We show that an increase in the
resistivity of dilute alloys at low temperatures is caused by by
an additional scattering of electrons by magnetic domain walls
which coincide with ferroelastic domain boundaries.

\begin{center}
---------------------------------------------------------------
\end{center}

[1] N.P.Armitage, D.H.Lu, C.Kim et al., Phys. Rev. Lett. \textbf{87}, 147003
(2001).

[2] A.Zimmers, Y.Noat, T.Cren, W.Sacks, D.Roditchev, B.Liang, and
R.L.Greene, Phys. Rev. B \textbf{76}, 132505 (2007).

[3] F.V.Prigara, arXiv:1106.5859 (2011).

[4] F.V.Prigara, arXiv:0708.1230 (2007).

[5] T.Iizuka, T.Mizuno, B.H.Min, Y.-S.Kwon, and S.Kimura, J. Phys. Soc. Jpn
(accepted), arXiv:1202.6169 (2012).

[6] F.V.Prigara, arXiv:1001.3061 (2010).

[7] S.Zhao, J.M.Mackie, D.E.MacLaughlin, O.O.Bernal, J.J.Ishikawa, Y.Ohta,
and S.Nakatsuji, Phys. Rev. B \textbf{83}, 180402(R) (2011).

[8] Y.Wang and N.P.Ong, Proc. Nat. Acad. Sci. USA \textbf{98}, 11091 (2001).

[9] T.Chatterji, G.J.McIntyre, and P.-A.Lindgard, Phys. Rev. B \textbf{79},
172403 (2009).

[10] J.Feinleib, in \textit{Electronic Structures in Solids}, edited by
E.D.Haidemenakis (Plenum Press, New York, 1969).

[11] G.I.Zhuravlev, \textit{Chemistry and Technology of Ferrites} (Khimiya
Publishers, Leningrad, 1970).

[12] R.Palai, R.S.Katiyar, H.Schmid et al., Phys. Rev. B \textbf{77}, 014110
(2008).

[13] D.Saint-James, G.Sarma, and E.J.Thomas, \textit{Type II
Superconductivity} (Pergamon Press, Oxford, 1969).

[14] J.S.Blakemore, \textit{Solid State Physics} (W.B.Saunders Company,
Philadelphia, 1969).

[15] R.J.Weiss, \textit{Solid State Physics for Metallurgists} (Pergamon
Press, Oxford, 1963).

[16] C.C.Schuler, in \textit{Optical Properties and Electronic Structure of
Metals and Alloys}, edited by F.Abeles (North-Holland Publishing Company,
Amsterdam, 1966).

[17] J.J.Hamlin, M.Janoschek, R.E.Baumbach, B.D.White, and M.B.Maple,
Philos. Mag. \textbf{92}, 647 (2012).

[18] F.V.Prigara, arXiv:1001.4152 (2010).

[19] K.Fischer, in \textit{Electronic Structures in Solids}, edited by
E.D.Haidemenakis (Plenum Press, New York, 1969).

[20] R.Okazaki, T.Shibauchi, H.J.Shi et al., Science \textbf{331}, 439
(2011).

[21] S.Fagot, P.Foury-Leylekian, S.Ravy, J.-P.Pouget, M.Anne, G.Popov,
M.V.Lobanov, and M.Greenblatt, Solid State Sci. \textbf{7}, 718 (2005).

\end{document}